\documentclass{article}
\usepackage[a4paper, total={6in,9in}]{geometry}
\usepackage[utf8]{inputenc}
\usepackage{graphicx}
\usepackage{authblk}
\usepackage[square,numbers]{natbib}
\usepackage{amsmath}

\begin{document}

\title{Single-shot terahertz spectrometer using a microbolometer camera}

\author{Dogeun Jang, Hanran Jin, and Ki-Yong Kim\textsuperscript{*}}

\affil{Institute for Research in Electronics and Applied Physics, University of Maryland, College Park, MD 20742}

\affil{*kykim@umd.edu}

\maketitle

\begin{abstract}
We demonstrate a single-shot terahertz spectrometer consisting of a modified Mach-Zehnder interferometer and a microbolometer focal plane array. The spectrometer is simple to use and can measure terahertz field autocorrelations and spectral power with no moving parts and no ultrashort-pulsed laser. It can effectively detect radiation at 10$\sim$40 THz when tested with a thermal source. It can be also used to measure the complex refractive index of a sample material. In principle, it can characterize both laser-based and non-laser-based terahertz sources and potentially cover 1$\sim$10 THz with specially-designed terahertz microbolometers.
\end{abstract}

%%%%%%%%%%%%%%%%%%%%%%%%%%  body  %%%%%%%%%%%%%%%%%%%%%%%%%%
\section{Introduction}
Terahertz (THz) spectroscopy is essential for many applications including material and biomedical science, defense and security, and imaging and communication\cite{grischkowsky1990far, ferguson2002materials, lee2009principles, mittleman2013sensing, kampfrath2013resonant}. Currently, THz time-domain spectroscopy (THz-TDS) is widely used for accurate THz field characterization with an electro-optic (EO) sampling  \cite{wu1995free, lee2009principles} or fast photoconductive switching \cite{auston1975picosecond,  lee2009principles} technique. %For accurate electric field characterizations, THz time-domain spectroscopy (THz-TDS) is widely used with electro-optic (EO) sampling \cite{wu1995free, lee2009principles}.
This method requires a multi-shot scan that takes from seconds to hours, depending
on the THz source power, repetition rate, desired scan step,
and field of view. Generally, this method is not well-suited for applications involving irreversible processes such as material damage and structural phase transitions or requiring real-time THz field characterization. Such applications greatly benefit from a single-shot technique.

So far various EO-based single-shot THz diagnostics have been developed \cite{jiang1998single, shan2000single, jamison2003high, kim2006single, kim2007single, teo2015invited, zheng2017common} and successfully applied, for example, to measure electron bunch lengths in real time \cite{wilke2002single, van2007single} and electrical conductivities of laser-ablated materials  \cite{kim2008measurements}. However, all EO-based techniques have fundamental limitations due to the dispersive and absorptive properties of OE materials, which can distort THz fields and restrict broadband detection \cite{nahata1996wideband, bakker1998distortion, gallot1999electro}. In the case of ZnTe, an OE crystal commonly used with 800 nm light sources, the detection bandwidth is typically limited to below 3 THz \cite{nahata1996wideband}. More importantly, all EO-based methods require ultrashort laser pulses for detection, limiting its applications to non-laser-based THz sources and systems.

As a stand-alone diagnostic, Fourier transform infrared (FTIR) spectroscopy \cite{griffiths2007fourier} is popularly used for THz or infrared characterization. Traditionally, FTIR spectroscopy uses a scanning scheme although there exist stationary (or single-shot capable) FTIR variants using detector arrays \cite{junttila1991performance}. A recent advent of focal plane array (FPA) sensors \cite{rogalski2011terahertz} that are sensitive to THz radiation  has enabled rapid and vibration-free THz characterization \cite{agladze2010terahertz}. In particular, microbolometer FPAs, typically used for thermal imaging at the long wavelength infrared (LWIR) region at 8$\sim$15 $\mu$m (or 20$\sim$37 THz), have been also applied for THz beam profiling and imaging at 1$\sim$20 THz. Recently, more advanced THz microbolometers have been developed by NEC \cite{oda2010uncooled} , INO \cite{bolduc2011noise}, and CEA Leti \cite{oden2013imaging} for applications at 1$\sim$10 THz.

In this article, we introduce a simple single-shot THz spectrometer using a microbolometer FPA. Like the EO-based single-shot techniques\cite{jiang1998single, shan2000single, jamison2003high, kim2006single, kim2007single, teo2015invited, zheng2017common}, our proposed scheme can characterize THz pulses in single shots. Also, as a FTIR spectrometer, it can provide additional advantages such as broadband THz detection and no necessity of an extra laser source. %Here we demonstrate single-shot THz characterization with a compact low-cost microbolometer camera, and then we examine its capability to be use THz spectroscopy by showing the examples of applications to the study of some materials.

\section{Experimental setup}

\begin{figure}[b!]
\centering\includegraphics{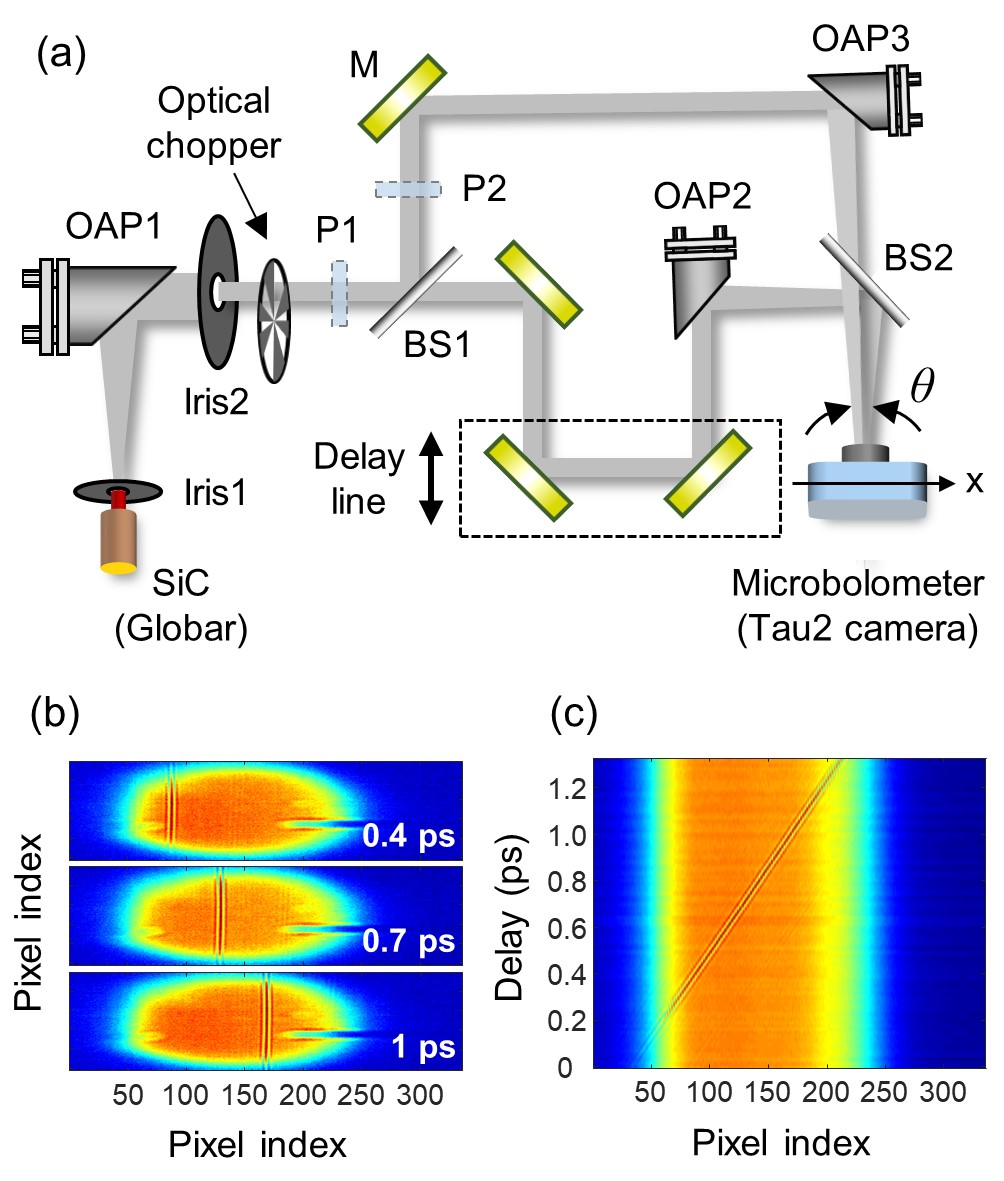}% Here is how to import EPS art
\caption{\label{fig1} (a) Schematic of single-shot THz detection using a FTIR interferometer. OAP = off-axis parabolic mirror; BS = beamsplitter; M = mirror. P = optional sample position. (b) Interferograms captured by the microbolometer at time delays of 0.4, 0.7, and 1 ps. (c) Line-outs from (b) patched and plotted as a function of the time delay ($y$-axis) between two beams .}
\end{figure}
Figure 1(a) shows a schematic diagram of our experimental setup to demonstrate a single-shot FTIR spectrometer. It consists of a modified Mach-Zehnder interferometer and uses a commercial vanadium oxide (VO$_x$) uncooled microbolometer camera (FLIR, Tau 2 336). The camera provides images of 336 $\times$ 256 pixels with one pixel size of 17 $\mu$m and can stream 14 bit images at a rate of 60 frames/sec with a camera link expansion board (FLIR) and a frame grabber (NI, PCIe-1433)\cite{oh2014generation, yoo2016generation, jang2019spectral}. To examine the THz spectrometer, a silicon carbide (SiC) blackbody radiator is used as a broadband THz source. The blackbody radiator has an active area of 3 mm $\times$ 4.4 mm with emissivity of 0.8, capable of reaching ~1700 K with 24 W electric power. Previously, this type of source was used to characterize the spectral response of the microbolometer over a broad range of THz frequencies \cite{jang2019spectral}. 

The radiation from the SiC emitter is collimated by an off-axis parabolic (OAP) mirror with the focal length of 4”, and its power and beam size are controlled by an iris diaphragm. The collimated THz beam is divided into two replica beams by a 1-mm-thick high-resistivity ($>$10 k$\Omega\cdot$cm) silicon (HR-Si) beamsplitter with the diameter of 2”. After reflection by gold mirrors, the two beams are  focused by separate OAP mirrors with the focal length of 7.5” and recombined by the second HR-Si beamsplitter. Finally, the two beams cross at an angle $\theta$ on the microbolometer camera, producing interference fringes. Here the parabolic mirrors (OAP2 and OAP3) are used to fit the beams into the microbolometer sensor size (5.7 mm $\times$ 4.4 mm). Also, to eliminate background thermal noise to the microbolometer, the beam from the SiC radiator is optically chopped at 5 Hz, and the image obtained with the beam blocked is subtracted from the one with the beam unblocked. Also the difference image is averaged over 30 frames to enhance the signal-to-noise ratio unless specified otherwise.

\section{Single-shot measurements}
In our crossing geometry, the relative time delay, $\tau$, between the two beams is mapped along the horizontal axis, $x$, [see Fig. 1(a)] on the microbolometer sensor surface as $\tau = 2x\text{sin}(\theta/2)/c$. Here $\theta$ is the crossing angle and $c$ is the speed of light in air. This relationship also determines the maximum possible time window as $T = 2w\text{sin}(\theta/2)/c$, where $w$ is the beam width. Also, the minimum temporal resolution, $\Delta \tau$, is determined by the single pixel size, $\Delta x$, as $\Delta \tau = 2\Delta x\text{sin}(\theta/2)/c$.

In the time domain, the interference between two THz fields, $E_1(t)$ and $E_2(t-\tau)$, can be expressed as 
\begin{equation}
\begin{split}
I(\tau) & = \int_{-\infty}^{+\infty} |E_{1}(t) + E_{2}(t-\tau)|^2dt,\\
& = I_{1} + I_{2} + \Delta I(\tau),
\label{eq1}
\end{split}
\end{equation}
where $I_1=\int_{-\infty}^{\infty}|E_1(t)|^2 dt$ and $I_2=\int_{-\infty}^{\infty}|E_2(t)|^2 dt$ are constants that do not depend on $\tau$. The last term $\Delta I(\tau)=\int_{-\infty}^{\infty}E_1(t)E_2^*(t-\tau) dt + \int_{-\infty}^{\infty}E_1^*(t)E_2(t-\tau) dt$ represents an cross-correlation function (or autocorrelation if $E_1 =E_2$). The Fourier transform of $\Delta I(\tau)$, 
\begin{equation}
\Delta \tilde{I}(\omega) = \tilde{E}_1(\omega)\tilde{E}_2^*(\omega) + \tilde{E}_1^*(\omega)\tilde{E}_2(\omega),
\label{eq2}
\end{equation}
provides information about the spectrum $\sqrt{I_1(\omega)I_2(\omega)}$.

In practice, the crossing angle, $\theta$, does not need to be measured directly. Instead, the mapping ratio $\tau / x$ can be measured by translating the delay line in Fig. 1(a) and performing a temporal scan. For example, Fig. 1(b) shows sample intreferograms taken at three different time delays $\tau$ = 0.4, 0.7, and 1 ps. Here $\tau$ = 0 is arbitrarily chosen when the interference fringes appear at the left edge of the beam. Figure 1(c) shows a series of interferogram line-outs (horizontal) patched together with varying time delay $\tau$. From the scan, $\tau/x \approx $7.4 fs/pixel is obtained, which provides $\theta = 7.5^{\circ}$.

\begin{figure}[t!]
\centering\includegraphics{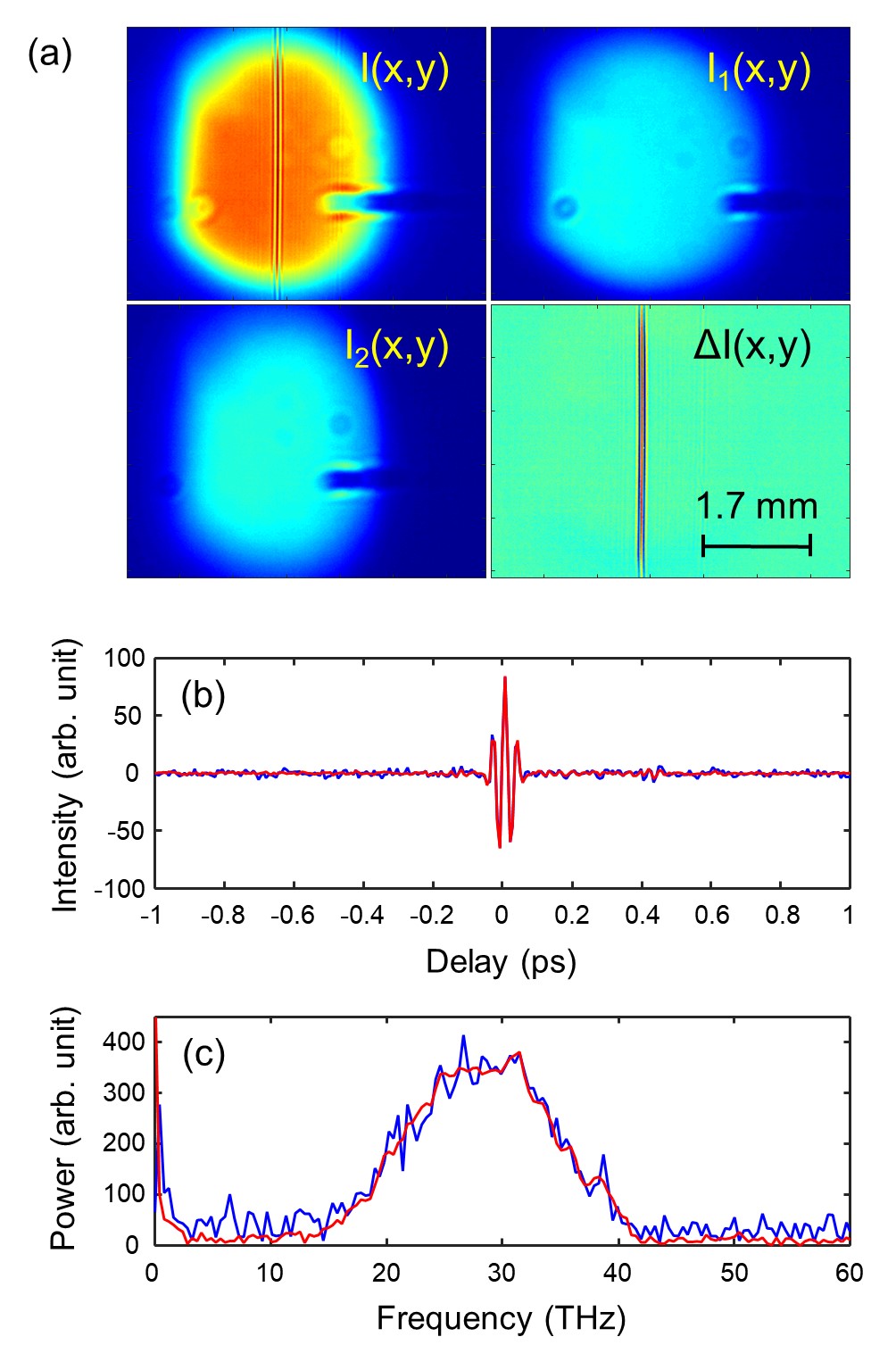}
\caption{\label{fig2} (a) Microbolometer images obtained with two beams on ($I$), one beam only ($I_1$, $I_2$), and the difference ($\Delta I = I-I_1-I_2$).  (b) THz autocorrelation obtained from a single (blue line) or 30-shot averaged (red line) $\Delta I$. (c) Corresponding THz spectra obtained via the Fourier transform.}
\end{figure}

Figure 2(a) shows THz beam profiles obtained with four different imaging modes.  $I(x,y)$ represents the intensity profile obtained with two interfering beams unblocked. $I_{1}(x, y)$ and $I_{2}(x, y)$ are the ones obtained with one beam unblocked while the other beam blocked. $\Delta I(x, y)$ is the difference intensity profile obtained from $\Delta I(x, y) =  I(x, y) - \{I_{1}(x, y) + I_{2}(x, y)\}$, which provides a background-free interferogram. Note that the horizontal stripe appearing in nearly all images in Fig. 1(b) and Fig. 2(a) is due to an optical damage made on the front silicon window of the microbolometer.

Figure 2(b) shows the THz autocorrelation, $\Delta I(\tau)$, obtained from $\Delta I(x, y)$ via space-to-time mapping, also with 100 line averaging along the $y$-axis. The time window obtained here is around 2.4 ps, which can provide a spectral resolution down to 0.42 THz. From the Fourier transform of the autocorrelation, $\Delta I(\tau)$, the corresponding power spectrum is computed and plotted in Fig. 2(c). For comparison, both 30-shot averaged and single-shot data are presented in Figs. 2(b) and 2(c), which are in good agreement. The measured spectrum, however, is far from blackbody radiation largely due to the microbolometer's strong response at the LWIR (20$\sim$37 THz) \cite{jang2019spectral}. In particular, the spectrum is strongly suppressed at $<$10 THz because of low beam power emitted from the SiC source and low sensitivity of the microbolometer at such frequencies. The spectrum also drops beyond 40 THz possibly due to a quarter-wave  optical  cavity adopted inside the microbolometer to limit the detection range \cite{jang2019spectral}. The enhanced signal at $<$1 THz is an artifact arising from our limited time window (2.4 ps).

\begin{figure}[t!]
\centering\includegraphics{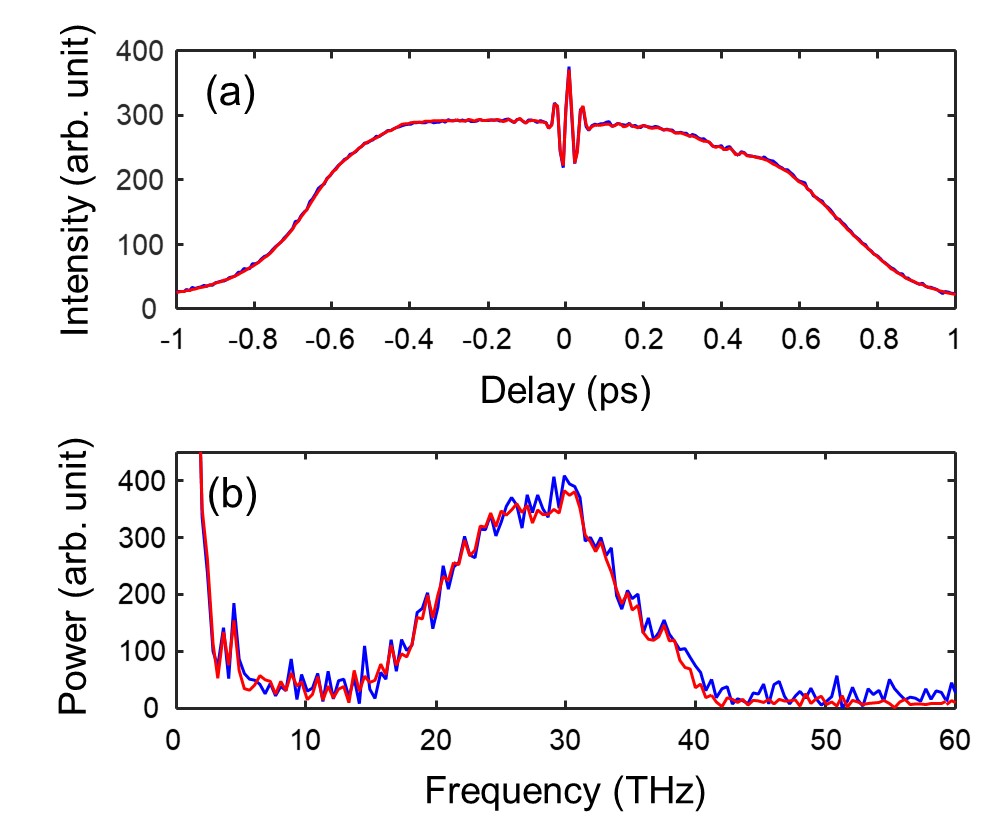}
\caption{\label{fig3} (a) THz autocorrelation obtained from $I(x,y)$ in Fig. 2(a). (b) Corresponding spectrum obtained via the Fourier transform. Red lines: 30-shot averaged. Blue lines: single shot.}
\end{figure}

An alternative method to retrieve the radiation spectrum is to use $I(x, y)$ only, not subtracted by $I_{1}(x, y)$ and $I_{2}(x, y)$. This method is applied to the data in Fig. 2(a), and the results are shown in Fig. 3. Notably, the autocorrelation in Fig. 3(a) exhibits a large slowly-varying DC offset due to the uncompensated background profiles $I_{1}(x, y)$ and $I_{2}(x, y)$. This, however, results in only low-frequency signals ($<$ 3 THz) in the spectrum due to the slow oscillation frequency associated with the beam intensity envelope. This near DC signal can be easily distinguished from the real spectrum. The measured spectrum here is well matched to the one in Fig. 2(c) except below 3 THz. Also, there is a good agreement between the 30-shot averaged (red lines) and single-shot (blue lines) data in Fig. 3. All these implies that a single-shot acquisition without separate beam profiling is possible if the radiation beam is sufficiently large with little intensity fluctuations, and the interested frequency is not close to DC.

\section{Transmission measurement}

Our single-shot THz spectrometer can be also utilized to characterize sample materials at THz frequencies. This can be done by placing a sample before the interferometer at P1 in Fig. 1(a). For example, Fig. 4 shows our measurements of THz transmission curves with three samples---a 280-$\mu$m-thick high-resistivity silicon (HR-Si) wafer, a 26-THz bandpass filter, and a 28-THz bandpass filter. Figures 4(a)(c)(e) show THz autocorrelations taken without (black lines, reference) and with (red lines) the sample. The corresponding spectra are obtained by the Fourier transform and plotted in Figs. 4(b)(d)(f). From the ratio of the power spectra, %obtained with to without the sample
the transmission curves are calculated and plotted (blue lines). Here the transmission is fairly well determined at 20$\sim$40 THz although it is quite noisy outside the range (blue-dotted line).% because of low THz power from the SiC source.

\begin{figure}[t!]
\centering\includegraphics{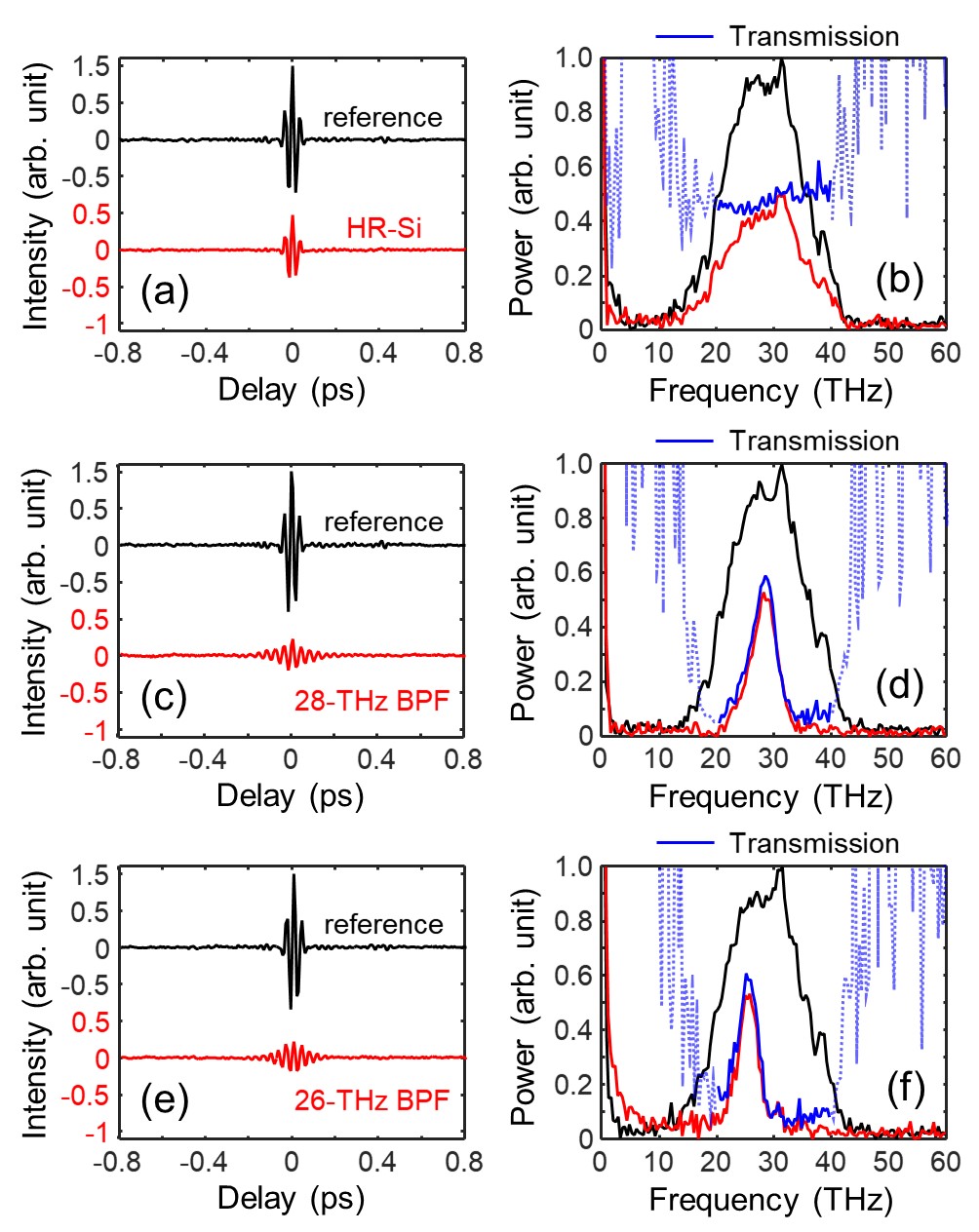}
\caption{\label{fig4} (a) THz autocorrelation with (red line) and without (black line) a Si placed in position P1 in Fig. 1(a). (b) Corresponding spectra (red and black lines), co-plotted with the resulting transmission curve of the sample (blue line). (c-d) Data measured with a 28-THz bandpass filter (BPF). (e-f) Data with a 26-THz BPF.}
\end{figure}
\section{Complex refractive index measurement}
As the second example, our single-shot spectrometer is used to extract the complex refractive index of $\tilde{n}\left(\omega\right) = n\left(\omega\right) + i\kappa\left(\omega\right)$ of a sample material. In this method, three autocorrelations must be obtained---without the sample,  $\Delta I(\tau)_{\text{ref}}$, with the sample placed in P1, $\Delta I(\tau)_{\text{P1}}$, and in P2, $\Delta I(\tau)_{\text{P2}}$ [see Fig. 1(a)]. For each case, the interfering THz fields can be expressed as
$\tilde{E}_1(\omega)=\tilde{E}_2(\omega)=\tilde{E}(\omega)$; $\tilde{E}_1(\omega)=\tilde{E}_2(\omega)=\sqrt{T(\omega)}\tilde{E}(\omega)e^{i \Delta \phi (\omega)}$; $\tilde{E}_1(\omega)=\tilde{E}(\omega)$, $\tilde{E}_2(\omega)=\sqrt{T(\omega)}\tilde{E}(\omega)e^{i \Delta \phi (\omega)}$, where  $\sqrt{T(\omega)}$  and $\Delta \phi (\omega)$ are the transmission amplitude and phase shift that the transmitting field acquires from the sample. From Eq. (\ref{eq2}), the Fourier transforms of the autocorrelation functions are given by
\begin{equation}
\begin{split}
\Delta \tilde{I}(\omega)_{\text{ref}} & = 2|\tilde{E}(\omega)|^2,\\
\Delta \tilde{I}(\omega)_{\text{P1}} & = 2T(\omega)|\tilde{E}(\omega)|^2,\\
\Delta \tilde{I}(\omega)_{\text{P2}} & = 2 \sqrt{T(\omega)}|\tilde{E}(\omega)|^2 \cos(\phi(\omega)).
\end{split}
\label{eq3}
\end{equation}
From Eq. (\ref{eq3}), the power transmission, $T(\omega)$ and phase shift, $\Delta \phi (\omega)$, can be retrieved as 
\begin{equation}
\begin{split}
T(\omega) & = \frac{\Delta \tilde{I}(\omega)_{\text{P1}}}{\Delta \tilde{I}(\omega)_{\text{ref}}},\\
\Delta \phi (\omega) & = \cos^{-1}\left(\frac{\Delta \tilde{I}(\omega)_{\text{P2}}}{\sqrt{\Delta \tilde{I}(\omega)_{\text{P1}}  \Delta \tilde{I}(\omega)_{\text{ref}} }} \right).
\end{split}
\label{eq4}
\end{equation}
Finally the complex refractive index of the sample with thickness $L$ can be obtained from $T(\omega)$ and $\Delta \phi (\omega)$ as %\cite{guo2007terahertz}

\begin{equation}
\begin{split}
n(\omega) &= 1 + \frac{c\Delta\phi(\omega)}{\omega L},\\ \kappa(\omega) &= -\frac{c}{\omega L}\ln\left(\frac{\left(n\left(\omega\right) + 1\right)^{2}}{4n(\omega)}\sqrt{T(\omega)}\right).
\end{split}
\label{eq5}
\end{equation}

\begin{figure}[t!]
\centering\includegraphics{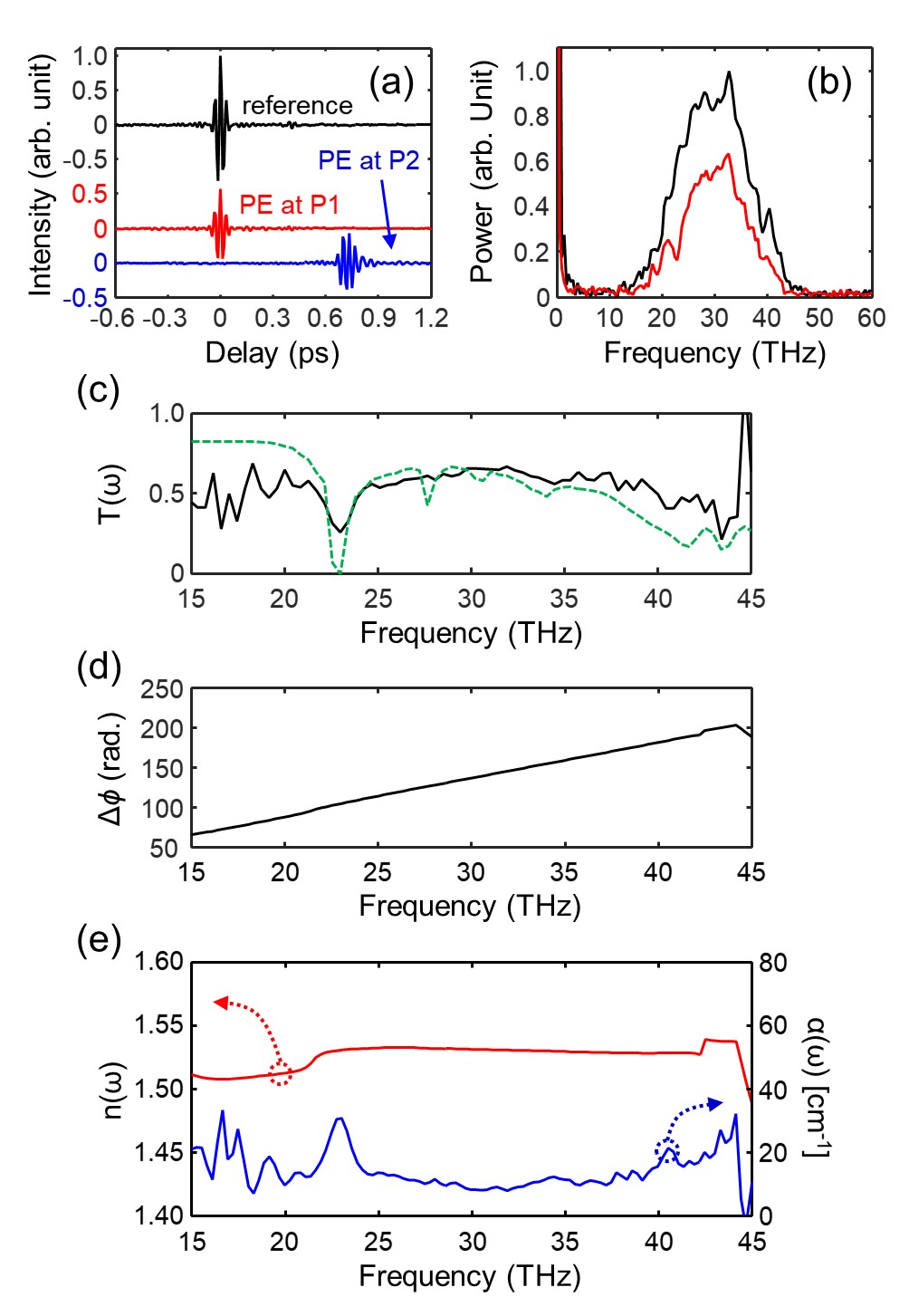}
\caption{\label{fig5} Demonstration of THz spectroscopy to characterize the complex refractive index of a 0.39-mm-thick polyethylene film. Measured (a) autocorrelations, (b) power spectra, (c) transmission curve (solid line), (d) phase shift, and (e) refractive index (red line) and absorption coefficient (blue line).}
\end{figure}

Figure 5 shows our measurement result obtained with a 390-$\mu$m-thick polyethylene (PE) film. Figure 5(a) shows three THz autocorrelations  $\Delta I(\tau)_{\text{ref}}$ (black line), $\Delta I(\tau)_{\text{P1}}$ (red line), and $\Delta I(\tau)_{\text{P2}}$ (blue line). The corresponding spectra and transmission curve are obtained from Eqs. (\ref{eq3}) and (\ref{eq4}) and plotted in Figs. 5(b) and 5(c). According to Fig. 5(c), the sample has $\sim$50\% transmission at 23$\sim$40 THz and exhibits a spectral dip at ~23 THz. The dip is caused by strong THz absorption by C-H bending vibration, also shown in the expected spectrum (green dotted line). Figure 5(d) shows the phase shift $\Delta \phi (\omega)$ extracted from Eq. (\ref{eq4}). Lastly, the refractive index, $n(\omega)$, and absorption coefficient, $\alpha(\omega) = 2\omega\kappa (\omega)/c$, are obtained from Eq. (\ref{eq5}) and plotted in Fig. 5(e). %It shows that PE film is ideal for a range for infrared applications with low absorption from 23-40 THz. An interesting feature of the refractive index (the red solid line) is that the PE film has a slightly negative dispersion at 23-40 THz.

\section{Conclusion}
In conclusion, we have demonstrated a simple single-shot FTIR spectrometer that combines a modified Mach-Zehnder interferometer with a microbolometer FPA. This can capture a THz field autocorrelation, equivalent to power spectrum, in a single shot with no moving parts.  %It is based on a FTIR setup with a modified Mach-Zehnder interferometer to form two crossing beams on the microbolometer FPA. uses crossing beam geometry with a microbolometer FPA.
It also allows spectral detection up to 40 THz, more than ten times higher than the typical EO-based single-shot methods \cite{jiang1998single, shan2000single, jamison2003high, kim2006single, kim2007single, teo2015invited, zheng2017common}. %about ten times higher than those achievable with typical EO-based single-shot methods.
Thus our diagnostic is best suited to characterize THz sources that emit high frequency radiation at $>$10 THz, operate at low repetition rates, or do not provide any synchronous optical light for EO characterization. Even at $<$10 THz, where our spectrometer has shown a poor response, %mainly due to the microbolometer's weak sensitivity at that region
our interferometric design can be still used with a THz microbolometer that is more sensitive at low THz frequencies \cite{oda2010uncooled, bolduc2011noise, oden2013imaging}. 

%The spectrometer demonstrated here has shown a poor response at $<$10 THz mainly due to our microbolometer's strong response at the LWIR region.  For characterization at <10 THz, specialized THz microbolometers that are more sensitive can be used.

\section*{Acknowledgments}
This work is supported by the Air Force Office of Scientific Research (FA9550- 25116-1-0163) and the Office of Naval Research (N00014-17-1-2705). The authors also thank Y. J. Yoo for his initial work on microbolometers.

\bibliographystyle{unsrt}
\bibliography{references}

\providecommand{\noopsort}[1]{}\providecommand{\singleletter}[1]{#1}%
\begin{thebibliography}{10}

\bibitem{grischkowsky1990far}
D~Grischkowsky, S{\o}ren Keiding, Martin Van~Exter, and Ch~Fattinger.
\newblock Far-infrared time-domain spectroscopy with terahertz beams of
  dielectrics and semiconductors.
\newblock {\em Journal of the Optical Society of America B}, 7(10):2006--2015,
  1990.

\bibitem{ferguson2002materials}
Bradley Ferguson and Xi-Cheng Zhang.
\newblock Materials for terahertz science and technology.
\newblock {\em Nature Materials}, 1(1):26--33, 2002.

\bibitem{lee2009principles}
Yun-Shik Lee.
\newblock {\em Principles of Terahertz Science and Technology}, volume 170.
\newblock Springer Science \& Business Media, 2009.

\bibitem{mittleman2013sensing}
Daniel Mittleman.
\newblock {\em Sensing with terahertz radiation}, volume~85.
\newblock Springer, 2013.

\bibitem{kampfrath2013resonant}
Tobias Kampfrath, Koichiro Tanaka, and Keith~A Nelson.
\newblock Resonant and nonresonant control over matter and light by intense
  terahertz transients.
\newblock {\em Nature Photonics}, 7(9):680, 2013.

\bibitem{wu1995free}
Qi~Wu and X-C Zhang.
\newblock Free-space electro-optic sampling of terahertz beams.
\newblock {\em Applied Physics Letters}, 67(24):3523--3525, 1995.

\bibitem{auston1975picosecond}
David~H Auston.
\newblock Picosecond optoelectronic switching and gating in silicon.
\newblock {\em Applied Physics Letters}, 26(3):101--103, 1975.

\bibitem{jiang1998single}
Zhiping Jiang and X-C Zhang.
\newblock Single-shot spatiotemporal terahertz field imaging.
\newblock {\em Optics Letters}, 23(14):1114--1116, 1998.

\bibitem{shan2000single}
Jie Shan, Aniruddha~S Weling, Ernst Knoesel, Ludwig Bartels, Mischa Bonn, Ajay
  Nahata, Georg~A Reider, and Tony~F Heinz.
\newblock Single-shot measurement of terahertz electromagnetic pulses by use of
  electro-optic sampling.
\newblock {\em Optics Letters}, 25(6):426--428, 2000.

\bibitem{jamison2003high}
Steven~P Jamison, Jingling Shen, Allan~M MacLeod, WA~Gillespie, and Dino~A
  Jaroszynski.
\newblock High-temporal-resolution, single-shot characterization of terahertz
  pulses.
\newblock {\em Optics Letters}, 28(18):1710--1712, 2003.

\bibitem{kim2006single}
K~Y Kim, B~Yellampalle, G~Rodriguez, R~D Averitt, A~J Taylor, and J~H Glownia.
\newblock Single-shot, interferometric, high-resolution, terahertz field
  diagnostic.
\newblock {\em Applied Physics Letters}, 88(4):041123, 2006.

\bibitem{kim2007single}
K~Y Kim, B~Yellampalle, A~J Taylor, G~Rodriguez, and J~H Glownia.
\newblock Single-shot terahertz pulse characterization via two-dimensional
  electro-optic imaging with dual echelons.
\newblock {\em Optics Letters}, 32(14):1968--1970, 2007.

\bibitem{teo2015invited}
Stephanie~M Teo, Benjamin~K Ofori-Okai, Christopher~A Werley, and Keith~A
  Nelson.
\newblock Invited article: Single-shot {TH}z detection techniques optimized for
  multidimensional {TH}z spectroscopy.
\newblock {\em Review of Scientific Instruments}, 86(5):051301, 2015.

\bibitem{zheng2017common}
Shuiqin Zheng, Xinjian Pan, Yi~Cai, Qinggang Lin, Ying Li, Shixiang Xu,
  Jingzhen Li, and Dianyuan Fan.
\newblock Common-path spectral interferometry for single-shot terahertz
  electro-optics detection.
\newblock {\em Optics Letters}, 42(21):4263--4266, 2017.

\bibitem{wilke2002single}
I~Wilke, Allan~M MacLeod, W~Allan Gillespie, G~Berden, GMH Knippels, and AFG
  Van Der~Meer.
\newblock Single-shot electron-beam bunch length measurements.
\newblock {\em Physical Review Letters}, 88(12):124801, 2002.

\bibitem{van2007single}
Jeroen van Tilborg, CB~Schroeder, Cs~T{\'o}th, CGR Geddes, Eric Esarey, and
  WP~Leemans.
\newblock Single-shot spatiotemporal measurements of high-field terahertz
  pulses.
\newblock {\em Optics Letters}, 32(3):313--315, 2007.

\bibitem{kim2008measurements}
K~Y Kim, B~Yellampalle, J~H Glownia, A~J Taylor, and G~Rodriguez.
\newblock Measurements of terahertz electrical conductivity of intense
  laser-heated dense aluminum plasmas.
\newblock {\em Physical Review Letters}, 100(13):135002, 2008.

\bibitem{nahata1996wideband}
Ajay Nahata, Aniruddha~S Weling, and Tony~F Heinz.
\newblock A wideband coherent terahertz spectroscopy system using optical
  rectification and electro-optic sampling.
\newblock {\em Applied Physics Letters}, 69(16):2321--2323, 1996.

\bibitem{bakker1998distortion}
H~J Bakker, G~C Cho, H~Kurz, Q~Wu, and X-C Zhang.
\newblock Distortion of terahertz pulses in electro-optic sampling.
\newblock {\em Journal of the Optical Society of America B}, 15(6):1795--1801,
  1998.

\bibitem{gallot1999electro}
Guilhem Gallot and D~Grischkowsky.
\newblock Electro-optic detection of terahertz radiation.
\newblock {\em Journal of the Optical Society of America B}, 16(8):1204--1212,
  1999.

\bibitem{griffiths2007fourier}
Peter~R Griffiths and James~A De~Haseth.
\newblock {\em Fourier transform infrared spectrometry}, volume 171.
\newblock John Wiley \& Sons, 2007.

\bibitem{junttila1991performance}
M-L Junttila, J~Kauppinen, and E~Ikonen.
\newblock Performance limits of stationary fourier spectrometers.
\newblock {\em Journal of the Optical Society of America A}, 8(9):1457--1462,
  1991.

\bibitem{rogalski2011terahertz}
A~Rogalski and F~Sizov.
\newblock Terahertz detectors and focal plane arrays.
\newblock {\em Opto-Electronics Review}, 19(3):346--404, 2011.

\bibitem{agladze2010terahertz}
Nikolay~I Agladze, J~Michael Klopf, Gwyn~P Williams, and Albert~J Sievers.
\newblock Terahertz spectroscopy with a holographic fourier transform
  spectrometer plus array detector using coherent synchrotron radiation.
\newblock {\em Applied Optics}, 49(17):3239--3244, 2010.

\bibitem{oda2010uncooled}
Naoki Oda.
\newblock Uncooled bolometer-type terahertz focal plane array and camera for
  real-time imaging.
\newblock {\em Comptes Rendus Physique}, 11(7-8):496--509, 2010.

\bibitem{bolduc2011noise}
Martin Bolduc, Marc Terroux, Bruno Tremblay, Linda Marchese, Eric Savard,
  Michel Doucet, Hassane Oulachgar, Christine Alain, Hubert Jerominek, and
  Alain Bergeron.
\newblock Noise-equivalent power characterization of an uncooled
  microbolometer-based {TH}z imaging camera.
\newblock In {\em Terahertz Physics, Devices, and Systems V: Advance
  Applications in Industry and Defense}, volume 8023, page 80230C.
  International Society for Optics and Photonics, 2011.

\bibitem{oden2013imaging}
Jonathan Oden, J{\'e}rome Meilhan, J{\'e}r{\'e}my Lalanne-Dera,
  Jean-Fran{\c{c}}ois Roux, Fr{\'e}d{\'e}ric Garet, Jean-Louis Coutaz, and
  Fran{\c{c}}ois Simoens.
\newblock Imaging of broadband terahertz beams using an array of
  antenna-coupled microbolometers operating at room temperature.
\newblock {\em Optics Express}, 21(4):4817--4825, 2013.

\bibitem{oh2014generation}
T~I Oh, Y~J Yoo, Y~S You, and K~Y Kim.
\newblock Generation of strong terahertz fields exceeding 8 {MV}/cm at 1 k{H}z
  and real-time beam profiling.
\newblock {\em Applied Physics Letters}, 105(4):041103, 2014.

\bibitem{yoo2016generation}
Yung-Jun Yoo, Donghoon Kuk, Zheqiang Zhong, and Ki-Yong Kim.
\newblock Generation and characterization of strong terahertz fields from k{H}z
  laser filamentation.
\newblock {\em IEEE Journal of Selected Topics in Quantum Electronics},
  23(4):1--7, 2016.

\bibitem{jang2019spectral}
Dogeun Jang, Malik Kimbrue, Yung-Jun Yoo, and Ki-Yong Kim.
\newblock Spectral characterization of a microbolometer focal plane array at
  terahertz frequencies.
\newblock {\em IEEE Transactions on Terahertz Science and Technology},
  9(2):150--154, 2019.

\end{thebibliography}
\end{document}